# CMOS-based Biosensor Arrays

R. Thewes, C. Paulus, M. Schienle, F. Hofmann, A. Frey, R. Brederlow, M. Augustyniak, M. Jenkner,
B. Eversmann, P. Schindler-Bauer, M. Atzesberger, B. Holzapfl, G. Beer, T. Haneder, and H.-C. Hanke

Infineon Technologies, Munich / Regensburg, Germany

## Abstract

*CMOS-based sensor array chips provide new and attractive features as compared to today's standard tools for medical, diagnostic, and biotechnical applications. Examples for molecule- and cell-based approaches and related circuit design issues are discussed.*

## 1. Introduction

Today's availability of highly integrated CMOS circuits on the one hand, and modern miniaturized biotech tools on the other, have had an essential impact on the way we live. In this paper, we consider the potential of merging both disciplines and related on-going developments in the fields of molecule- and cell-based applications.

As an example a simplified flow chart scenario of the drug development process is schematically depicted in Fig. 1. Moreover, CMOS-based solutions also provide a high potential for diagnostic applications.

## 2. DNA Microarrays

The purpose of DNA microarray chips [1-3] is the parallel investigation concerning the amount of specific DNA sequences in a given sample. Within predefined positions, single-stranded DNA receptor (probe) molecules are immobilized on the surface of such chips (Fig. 2a)). For application, the whole chip is flooded with a sample containing the target molecules. Complementary (i.e. matching) sequences of probe and target molecules hybridize (Fig. 2d)), whereas that chemical binding process does not occur in case of mismatching strands (Fig. 2e)). Finally, after a washing step, double-stranded DNA is obtained at the match positions (Fig. 2f)), and single-stranded DNA at the mismatch sites (Fig. 2g)). Identification of the sites with double-stranded DNA thus reveals the composition of the sample, since the probes and their positions are known.

Whereas optical detection principles make use of fluorescence or chemoluminescense light originating from label molecules bound to the targets [1-3], electronic principles utilize electrochemically active labels or labels which trigger electrochemical activity to translate hybridization events into charge or current collected by (noble metal) sensor electrodes [4-6]. Alternative label-free principles are under development. They focus on the effect of impedance or mass changes at the sensors' surfaces after hybridization [7-11].

Using a redox-cycling based technique, CMOS chips have recently been published which detect currents between 1pA and 100nA per sensor [12, 13]. There, within each sensor site the sensor signal is A/D-converted using a current-to-frequency converting sawtooth generator concept. As shown in Fig. 3, the voltage of the sensor electrode is controlled by a regulation loop via an operational amplifier and a source follower transistor. An integrating capacitor $C_{int}$ is charged by the sensor current. When the switching level of the comparator is reached, a reset pulse is generated. The measured frequency is approximately proportional to the sensor current. For A/D conversion, the number of reset pulses is counted with a digital counter within a given time frame.

A chip photo of the latest version of such chips using the above described principle is shown in Fig. 4. The chips consist of a 8×16 sensor array including peripheral circuitry (bandgap and current references, auto-calibration circuits, D/A-converters to provide the required voltages for the electrochemical operation) and 6 pin interface for power supply and serial digital data transmission.

Measured electrical, electrochemical, and biological data as well as processing related issues are published elsewhere [12-15].

## 3. Recording from nerve cells and neural tissue

The elementary neural signals of cells, action potentials, are temporal peaks of the intracellular voltage, which are associated with ion currents through the cell membrane. When neurons within a electrolyte are brought in intimate contact with a planar surface, a cleft of order of 60nm between cell membrane and surface is obtained. Ion currents flowing through the cleft lead to a potential drop due to the resistance of the cleft, which can be capacitively probed with an Oxide-Semiconductor-FET (OSFET) below the cleft [16-18].

Fig. 5 depicts a CMOS-compatible approach enabling high spatial resolution: The cleft voltage is capacitively coupled to an electrode covered by a thin dielectric and connected to the gate of a standard MOSFET. On this basis, chips with 128×128 positions within a total sensor area of 1mm×1mm are presented in [19]. Since typical neuron diameters are 10µm…100µm the chosen pitch of 7.8µm guarantees that each cell is monitored independent of its individual position. Full frame rate is 2k samples/s.

Circuit design related details are shown in Fig. 6. Selection of a pixel within a row of the array is performed by closing the related switch $S_2$. Since the maximum signal amplitudes are between 100µV and 5mV, the sensor MOSFETs ($M_1$) must be calibrated to compensate for the effect of their parameter variations. This is done by closing switch $S_1$ and forcing a current through $M_1$ by current source $M_2$. After opening $S_1$ again, a voltage related to the calibration current is stored on the gate of $M_1$. This process is periodically performed for all rows in parallel and for all columns in sequence, so that all sensor transistors $M_1$ within a row provide the same current when selected independent of their individual device parameters.

In the readout mode $S_1$ is open, $S_3$ is closed, and $M_2$ is operated with the same current as in the calibration mode. Action potential induced changes of the sensor electrode potential lead to difference currents between $M_1$ and $M_2$ which are compensated by the closed regulation loop composed of A, $M_3$, and $M_4$ and further amplified through the whole signal path shown in Fig. 6. As can be seen, the subsequent current gain stages also undergo a calibration procedure before used for signal amplification.

Measured electrical and biological results, and processing related issues are published elsewhere [20, 21].

## 4. Summary

CMOS-based biosensor arrays clearly have proven their feasibility, they open the way to novel and user-friendly highly parallel solutions. One of the remaining challenges is to make commercialization a successful story.

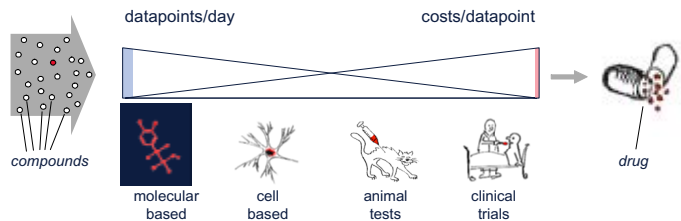

Fig. 1: Schematic diagram depicting the drug-screening process flow aiming to identify one (combination of) compound(s) out of millions of (combinations of) compounds from a library as a suitable drug for a given purpose.

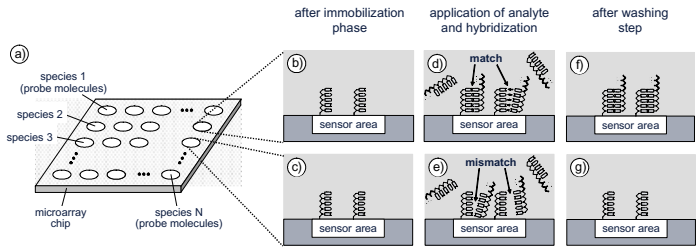

Fig. 2: a) Schematic plot of a DNA microarray chip. b), c): Schematic consideration of two different test sites after immobilization process. For simplicity, probe molecules with five bases only are shown here (real applications: typically 15–40). d), e): Hybridization phase. A sample containing target molecules to be detected (up to 2…3 orders of magnitude longer compared to the probe molecules) is applied to the whole chip. Hybridization occurs in case of matching DNA strands (d). In case of mismatching molecules (e), chemical binding does not occur. f), g): Situation after subsequent washing step.

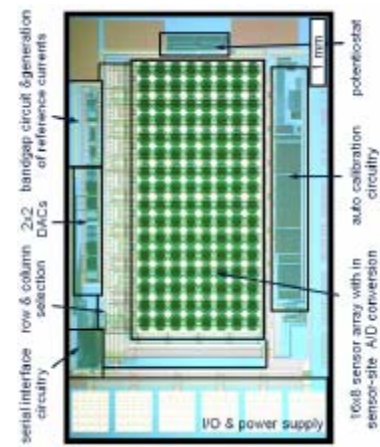

Fig. 4: Chip photo of a CMOS-based 16×8 DNA microarray [15]. Basic CMOS Process: $L_{min}=0.5\mu m$, $t_{ox}=15nm$, $V_{DD}=5V$.

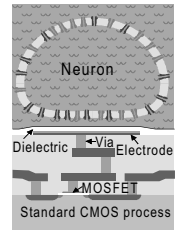

Fig. 5: Schematic cross section of a CMOS-compatible sensor approach for non-invasive neural signal recording [19-21].

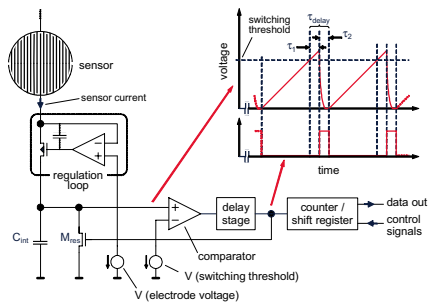

Fig. 3: Circuit principle used for in-sensor site A/D-conversion of the sensor signal based on current-to-frequency conversion [13].

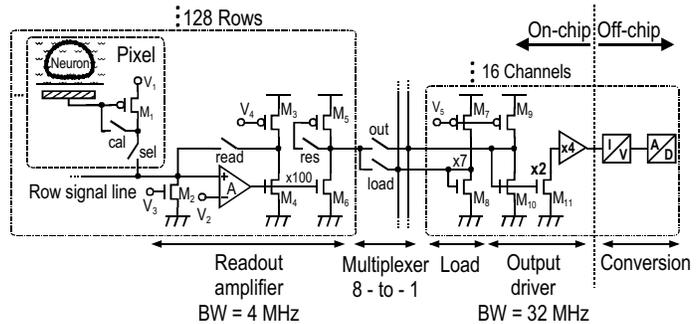

Fig. 6: Complete signal path of the sensor chip presented in [19, 20] and used in [21].